\documentclass[a4paper,twoside,10pt]{article}
\usepackage{graphicx,publaob}

\def\papertype{Contributed paper}
\begin{document}

\title{Rotation of 10 B\lowercase{e} stars through Fourier transform analysis}

\authors{Nata\v sa Gavrilovi\' c}

\address{$^1$Astronomical Observatory, Volgina 7, 11000 Belgrade, Yugoslavia}
\Email{ngavrilovic}{aob.aob.bg.ac}{yu}

\markboth{Rotation of 10 B\lowercase{e} stars through Fourier transform analysis}{N. Gavrilovi\' c}

\abstract{Here we determine the projected rotational velocity of 10 Be stars using Fourier Transform Method. Also, we discuss the gravity darkening and extend of deviation from solid body rotation for our sample of stars. We found that 7 of considered stars are affected by strong gravity darkening or/and  solar differential rotation.}
\section{INTRODUCTION}

Since Be stars are fast rotators, our knowledge about their fundamental parameters, in particular the projected rotational velocity, is subject to considerable uncertainty. Their gravity and temperature are aspect angle-dependent and do not have the straightforward meaning as they have in slow rotators (Slettebak 1976,1982; Brown \& Verschneren 1997; Brown \& Verschueren 1997; Halbedel 1996).

  Projected rotational velocity $V_{e}\sin(i)$ for ten  stars in our sample of Be stars has been determined using spectra from CDS data base\footnote{http://cdsweb.u-strasbg.fr/cgi-bin/Cat?J/A\%2bA/378/861}.
In order to determine the projected rotational velocity, Fourier Transform Method-FTM (Jankov 1995; Jankov et al. 2000) has been applied to each individual spectrum of our sample of stars, since the position of the first minimum is determined by the rotational broadening, and does not depend on other broadening mechanism. Reiners (2003) showed that the ratio of the first two minimum positions $q_{2}/q_{1}$ of the Fourier Transform can be also a reliable parameter to conclude the amount of gravity darkening or differential rotation. A value of $q_{2}/q_{1}< 1.72$ is a direct indication for a  gravity darkening or a solar-like differential rotation law, while $q_{2}/q_{1}> 1.83$ indicates anti-solar  differential rotation. The measurement of $q_{2}/q_{1}$ can be used without any modeling of line profiles.

\begin{table}
\begin{center}
\begin{tabular}{|c|c|c|c|c|}
\hline
$HD$    &  $Wavelenght[nm]$ & $V_{e}\sin(i)[km/s]$ & $q_{2}/q_{1}$ & $V_{e}\sin(i)[km/s]^{*}$\\
\hline
10144 &    HeI 438.81     &    217 & 1.62   & 235  \\
56139 &      HeI 438.81  &    88 &  1.16 &  85 \\
57219 &      HeI 438.81 &    90 & 1.17  & 80 \\
75311 &      HeI 438.81 &    264 & 1.74  & 268 \\
77320 &      HeI 438.81 &    338 & 1.49 & 345 \\
189687&      HeI 438.81 &  211& 1.27  & 200 \\
201733&     HeI 438.81&  351  & 2.22 & 340 \\
210129&    HeI 438.81 &  141  & 1.70 & 130 \\
120324&    HeI 438.81 &  156  & 1.74 & 124 \\
224544&   HeI 438.81 &  239  & 1.36 & 260 \\
\hline
\end{tabular}
\caption{Determined values of $V_{e}\sin(i)$ in [km/s] and $q_{2}/q_{1}$ for the sample of 10 Be
 stars. In the last column (denoted with *) are values given by Chauville et al.(2001)}
\end{center}
\end{table}

\begin{figure}
\begin{center}
\includegraphics[width=9cm,height=11cm,angle=270]{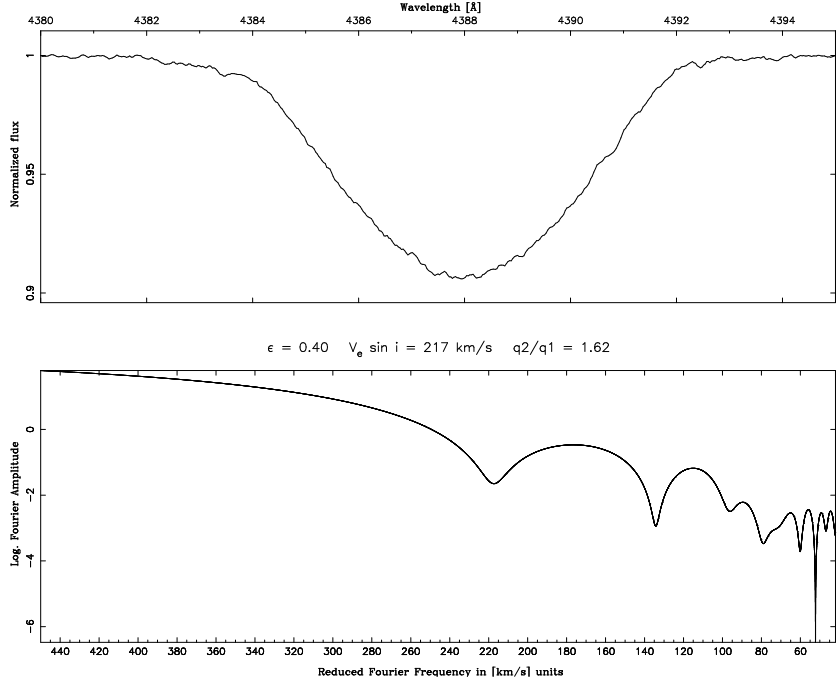}
\caption{ Flux line profile of the HeI 438.8 nm line of HD 10144 (top) and its Fourier transform (bottom). The Fourier frequency was reduced to velocity units, so that the first minimum of the Fourier transform of a rotational profile points to the projected rotational velocity of the star.}
\end{center}
\end{figure}

\section{RESULTS AND DISCUSSION}

 Using FTM and values for limb-darkening coefficient given by Claret (2000), we obtained projected rotational velocity, and also ratio of the first two minimum positions for 10 Be stars in our sample.  In Table 1. are given values for $q_{2}/q_{1}$, $V_{e}\sin(i)$ and (the last column) the determination of $V_{e}\sin(i)$ for considered stars by Chauville et al. (2001). As one can see from Table 1, our calculated values for $V_{e}\sin(i)$ are in good agreement with those determined by Chauville et al. (2001). Only three stars from the group have value $q_{2}/q_{1}>1.72$, so it can be concluded that the stars HD 75311 and HD 120324 have solid-body rotation and the mass concentration in the center of the star. The third one, HD 201733 has anti-solar differential rotation. 
Our analysis for the rest of the stars in the sample revealed a small $q_{2}/q_{1}$ ratio (less than 1.72) which could be interpreted as gravity darkening or/and solar differential rotation.\\
 In the case of Achernar (HD 10144), no evidence has been found for differential rotation in high quality spectroscopy of the star (e.g. Gray 1977; Howarth \& Smit  2001). It means that our result implies a strong gravity darkening which could be expected taking into account the extremely oblate shape of the star (Domiciano de Souza et al. 2003).

\references
Brown, A.G.A., \& Verschueren, W. 1997, \journal{Astron. Astrophys.}, \vol{319}, 811

Chauville, J., Zorec, J., Ballereau, D. et al. 2001, \journal{Astron. Astrophys.}, \vol{378}, 861 (Tables are only available in full in electronic form at CDS via\\ 
http://vizier.u-strasbg.fr/viz-bin/VizieR

Claret, A. 2000,\journal{Astron. Astrophys.},\vol{363}, 1081

Domiciano de Souza, A., Kervella, P., Jankov, S.,Abe, L., Vakili, F., et al. 2003, \journal{Astron. Astrophys.}, \vol{407L}, 47

Gray, D. F. 1977 \journal{ApJ}, \vol{211}, 198

Halbedel, E.M. 1996, \journal{PASP}, \vol{108}, 833

Howarth, I.D., Smith, K.C., 2001 \journal{MNRAS}, \vol{327}, 353

Jankov S.,1995, Publ. Obs. Astron.Belgrade, \vol{50}, 75  

Jankov S., Janot-Pacheco, E., Leister, N.V., 2000, \journal{ApJ}, \vol{540}, 535

Reiners, A. 2003, \journal{Astron. Astrophys.}, \vol{408}, 707

Slettebak, A., Collins II, G.W., \& Traux, R. 1992, \journal{ApJS}, \vol{81}, 335

Steele, I.A., Negueruela, I., \& Clark, J.S. 1999, \journal{A \& AS}, \vol{137}, 147

\endreferences

\end{document}